\documentclass[12pt,aps,showpacs]{revtex4}%
\usepackage{amssymb}
\usepackage{amsmath}
\usepackage{amsfonts}
\usepackage{graphicx}%
\providecommand{\U}[1]{\protect\rule{.1in}{.1in}}
\begin{document}
\title{Killing Tensors and Symmetries}
\author{David Garfinkle}
\affiliation{Physics Department, Oakland University, Rochester, MI}
\author{E.N. Glass}
\affiliation{Physics Department, University of Michigan, Ann Arbor, MI}
\date{\today}

\begin{abstract}
A new method is presented for finding Killing tensors in spacetimes with
symmetries. The method is used to find all the Killing tensors of Melvin's
magnetic universe and the Schwarzschild vacuum.\ We show that they are all
trivial. The method requires less computation than solving the full Killing
tensor equations directly, and it can be used even when the spacetime is not
algebraically special

\end{abstract}

\maketitle

\section{Introduction}

Killing tensors are useful because, like Killing vectors, they provide
conserved quantities for geodesic motion, most famously in the Kerr metric
where the Killing tensor gives rise to the Carter constant \cite{Car68}.
However, it is much more difficult to find Killing tensors than Killing
vectors. In 4 spacetime dimensions, the equation for a Killing tensor becomes
20 partial differential equations for 10 functions of 4 variables. Most of the
known results on Killing tensors come from the fact that the Killing tensor
equation simplifies in certain classes of algebraically special spacetimes
\cite{Som73,KSH+03}. However, equations in general relativity also often
simplify in the presence of symmetry \cite{Ger72}. We will show that the
Killing tensor equation simplifies when a spacetime possesses a hypersurface
orthogonal Killing vector, and that this simplification provides an effective
method for finding Killing tensors. We will apply this method to Melvin's
magnetic universe \cite{Mel64} and the Schwarzschild vacuum solution. It is
shown that all of their Killing tensors are trivial in the sense that they are
either the metric or the symmetrized product of Killing vectors. The method is
presented in section 2. It is applied to Melvin's magnetic universe in section
3 and the Schwarzschild vacuum in section 4. Conclusions are given in section 5.

\ \newline\textbf{Notation}: Lower case Latin indices, $B^{a}$, range over
n-dimensions. Greek indices, $B^{\mu}$, range over n--1 dimensions. For
Killing vector $\xi^{a}$ an overdot will denote a Lie derivative, $\dot
{A}:=\mathcal{L}_{\xi}A.$

\section{The Killing tensor method}

A Killing tensor (of order 2) is a symmetric tensor $X_{ab}$ that satisfies
\begin{equation}
{\nabla_{(a}}{X_{bc)}}=0
\end{equation}

Now suppose that the spacetime has a hypersurface orthogonal Killing vector
$\xi^{a}$. Define $V$ such that
\begin{equation}
{\xi^{a}}{\xi_{a}}=\epsilon{V^{2}}%
\end{equation}
where $\epsilon=\pm1$. Then the metric in directions orthogonal to $\xi^{a}$
is given by
\begin{equation}
{h_{ab}}={g_{ab}}-\epsilon{V^{-2}}{\xi_{a}}{\xi_{b}}%
\end{equation}
One can use ${h^{a}}_{b}$ as a projection operator to project any tensor in
directions orthogonal to $\xi^{a}$. In particular, the Killing tensor can be
decomposed as
\begin{equation}
{X_{ab}}=A{\xi_{a}}{\xi_{b}}+2{B_{(a}}{\xi_{b)}}+{C_{ab}}
\label{decomposition}%
\end{equation}
where $B_{a}$ and $C_{ab}$ are orthogonal to $\xi^{a}$. The Killing tensor
equation for this decomposition is
\begin{equation}
\nabla_{(a}A{\xi_{b}}{\xi_{c)}}+2\nabla_{(a}{B}_{b}{\xi_{c)}}+\nabla
_{(a}{C_{bc)}=0.}%
\end{equation}
Projecting the Killing tensor equation using all combinations of ${h^{a}}_{b}$
and $\xi^{a}$ yields the following
\begin{align}
{D_{(a}}{C_{bc)}}  &  =0,\label{kt1}\\
{\mathcal{L}_{\xi}}A  &  =-2{V^{-1}}{B^{a}}{D_{a}}V,\label{kt2}\\
{\mathcal{L}_{\xi}}{B_{a}}  &  =-{V^{-1}}{C_{ab}}{D^{b}}V-{{\frac{1}{2}}%
}\epsilon{V^{2}}{D_{a}}A,\label{kt3}\\
{\mathcal{L}_{\xi}}{C_{ab}}  &  =-2\epsilon{V^{2}}{D_{(a}}{B_{b)}.}
\label{kt4}%
\end{align}
Here $D_{a}$ is the derivative operator associated with the metric $h_{ab}$
and $\mathcal{L}_{\xi}$ denotes the Lie derivative with respect to Killing
vector $\xi^{a}$. Solving Eq.(\ref{kt3}) for ${D_{a}}A$ yields
\begin{equation}
{D_{a}}A=-(2/\epsilon)\left[  {V^{-2}}{\mathcal{L}_{\xi}}{B_{a}}+{V^{-3}%
}{C_{ab}}{D^{b}}V\right]  \label{spatialgrada}%
\end{equation}
The left hand side of Eq.(\ref{spatialgrada}) is curl-free and thus the curl
of the right hand side must vanish. Hence
\begin{equation}
{D_{[a}}({V^{-2}}{\mathcal{L}_{\xi}}{B_{b]}})+{D_{[a}}({V^{-3}}{C_{b]c}}%
{D^{c}}V)=0 \label{ic1}%
\end{equation}
Furthermore, since $\mathcal{L}_{\xi}$ commutes with $D_{a}$ it follows that
$\mathcal{L}_{\xi}$ of the right hand side of Eq.(\ref{spatialgrada}) equals
$D_{a}$ of the right hand side of Eq.(\ref{kt2}), and therefore
\begin{equation}
{\mathcal{L}_{\xi}}{\mathcal{L}_{\xi}}{B_{a}}+{V^{-1}}({D^{b}}V){\mathcal{L}%
_{\xi}}{C_{ab}}-\epsilon{V^{2}}{D_{a}}({V^{-1}}{B^{b}}{D_{b}}V)=0. \label{ic2}%
\end{equation}
Equations (\ref{ic1}) and (\ref{ic2}) provide the integrability conditions for
equations (\ref{kt2}) and (\ref{spatialgrada}).

These equations are most easily implemented in a coordinate system adapted to
the Killing vector. Choose a coordinate system ($y,{x^{\mu}}$) such that
$x^{\mu}$ are coordinates on the surface orthogonal to the Killing vector and
$\mathcal{L}_{\xi}$ is simply a partial derivative with respect to $y$. Use
$\partial_{\mu}$ or a comma to denote a derivative with respect to the
$x^{\mu}$ coordinates. The Latin indices in this section are n-dimensional,
and the method below projects objects and equations down to n-1 dimensions
with Greek indices.

Equation (\ref{kt4}) becomes
\begin{equation}
{{\dot{C}}_{\mu\nu}}=-\epsilon{V^{2}}({B^{\alpha}}{\partial_{\alpha}}%
{h_{\mu\nu}}+{h_{\alpha\nu}}{\partial_{\mu}}{B^{\alpha}}+{h_{\mu\alpha}%
}{\partial_{\nu}}{B^{\alpha}}) \label{kt4a}%
\end{equation}
while integrability conditions (\ref{ic1}) and (\ref{ic2}) become
\begin{align}
{\partial_{\lbrack\mu}}({V^{-2}}{{\dot{B}}_{\nu]}})+{\partial_{\lbrack\mu}%
}({V^{-3}}{C_{\nu]\alpha}}{\partial^{\alpha}}V)  &  =0\label{ic1a}\\
{{\ddot{B}}_{\mu}}+{V^{-1}}{{\dot{C}}_{\mu\nu}}{\partial^{\nu}}V-\epsilon
{V^{2}}{\partial_{\mu}}({V^{-1}}{B^{\nu}}{\partial_{\nu}}V)  &  =0.
\label{ic2a}%
\end{align}
Equations (\ref{kt2}) and ({\ref{spatialgrada}) for $A$ are written as
\begin{align}
{\dot{A}}  &  =-2{V^{-1}}{B^{\mu}}{\partial_{\mu}}V\label{kt2a}\\
{\partial_{\mu}}A  &  =-(2/\epsilon)({V^{-2}}{{\dot{B}}_{\mu}}+{V^{-3}}%
{C_{\mu\nu}}{\partial^{\nu}}V) \label{spatialda}%
\end{align}
Note that equations (\ref{kt4a}-\ref{spatialda}) can be evaluated without ever
having to calculate a Christoffel symbol. }

The complete set of Killing tensors can be found as follows: first solve
Eq.(\ref{kt1}) for the most general $C_{\mu\nu}$. Note that this general
solution will contain arbitrary \textquotedblleft constants\textquotedblright%
\ that are really functions of the Killing coordinate. Now using the general
$C_{\mu\nu}$, find the most general $B_{\mu}$ satisfying Eq.(\ref{kt4a}). Then
restrict this general solution by demanding that equations (\ref{ic1a}) and
(\ref{ic2a}) also be satisfied. Finally, solve equations (\ref{kt2a}) and
(\ref{spatialda}) for $A$.

\section{Killing tensors of the Melvin metric}

Melvin's magnetic universe is a static, cylindrically symmetric, Petrov type
\textbf{D} solution of the Einstein-Maxwell equations. Its metric is usually
written as
\begin{equation}
d{{\tilde{s}}_{\text{Mel}}^{2}}={a^{2}}(-d{{\tilde{t}}^{2}}+d{\rho^{2}%
}+d{{\tilde{z}}^{2}})+(\rho^{2}/a^{2})d{{\tilde{\phi}}^{2}}%
\end{equation}
where the function $a$ is
\begin{equation}
a=1+{{\frac{1}{4}}}{B_{0}^{2}}{\rho^{2}.}%
\end{equation}
The constant $B_{0}$ is the value of the magnetic field on the $\rho=0$ axis.
Define coordinates $t=({B_{0}/2)\tilde{t}},$ $r=({B_{0}/2)}\rho,\,z=({B_{0}%
/2)\tilde{z}},\,\phi={\tilde{\phi}}$. In terms of these coordinates we have
\begin{equation}
a=1+{r^{2}}%
\end{equation}
while the metric becomes $d{{\tilde{s}}_{\text{Mel}}^{2}}=(4/{B_{0}^{2}%
})d{s_{\text{Mel}}^{2}}$ with
\begin{equation}
d{s_{\text{Mel}}^{2}}={a^{2}}(-d{t^{2}}+d{r^{2}}+d{z^{2}})+(r^{2}/a^{2}%
)d{\phi^{2}.}%
\end{equation}
Since $d{{\tilde{s}}_{\text{Mel}}^{2}}$ and $d{s_{\text{Mel}}^{2}}$ differ
only by an overall constant scale, they have the same Killing vectors and
Killing tensors. For simplicity, we will work with metric $d{s_{\text{Mel}%
}^{2}}$. Since the metric components are independent of $t,\,\phi,$ and $z,$
it follows that there are Killing vectors for each of these coordinate
directions. They are denoted by $(t,\phi,z)\rightarrow(\tau^{\mu},\eta^{\mu
},\lambda^{\mu})$ respectively. Each of these Killing vectors is hypersurface
orthogonal. In addition, the metric has boost symmetry in the $tz$ plane, with
corresponding Killing field $t{\lambda^{a}}+z{\tau^{a}}$.

We will use the method of the previous section to work out the Killing tensors
of Melvin's magnetic universe. First we will find the Killing tensors of the
2-dimensional $rz$ surface, then use these to find the Killing tensors of the
3-dimensional $rz\phi$ surface, and finally find the Killing tensors of the
4-dimensional Melvin metric.

We will use ${c_{1}},\,{c_{2}}$ etc. to denote constants, and ${k_{1}%
},\,{k_{2}}$ etc. to denote quantities that depend only on the coordinate
associated with the Killing vector.

\subsection{An $rz$ surface}

The $rz$ 2-surface has a metric of form
\begin{equation}
{g}_{ab}dx^{a}dx^{b}={a^{2}}(d{r^{2}}+d{z^{2}}).
\end{equation}
We have a $z$ coordinate Killing vector $\lambda^{\mu}$ for which $V=a$ and
$\epsilon=1$, and a 1-dimensional metric ${h_{\mu\nu}}={a^{2}}r,_{\mu}r,_{\nu
}$ orthogonal to the Killing vector. Since the r-direction is a 1-dimensional
line, it follows that the Killing tensor must take the form $C_{\mu\nu
}=F{h_{\mu\nu}}$ for some scalar $F$. It then follows from Eq.(\ref{kt1}) that
$F$ is independent of $r$. We therefore have
\begin{equation}
{C_{\mu\nu}}={k_{1}h_{\mu\nu}=k}_{1}{a^{2}}r,_{\mu}r,_{\nu}%
\end{equation}
for some $k_{1}(z)$. Equation (\ref{kt4a}) then becomes, with $B^{\mu
}\rightarrow B^{r}$ only
\begin{equation}
{{\dot{C}}_{rr}}={{\dot{k}}_{1}a}^{2}=-a^{2}(\frac{da^{2}}{dr\ }B^{r}%
+2a^{2}\partial_{r}B^{r})
\end{equation}
or%
\begin{equation}
{{\dot{k}}_{1}}=-2{a^{2}}\left(  \frac{d{B^{r}}}{dr}+{\frac{1}{a}}{\frac
{da}{dr}}{B^{r}}\right)
\end{equation}
The general solution for $B^{r}$ is
\begin{equation}
{B^{r}}={a^{-1}}\left(  {k_{2}}-{\frac{{\dot{k}}_{1}}{2}}\text{arc}{\tan
\ }r\right)
\end{equation}
Equation (\ref{ic1a}) is automatically satisfied, while Eq.(\ref{ic2a})
becomes
\begin{equation}
0=(-{{\frac{1}{2}}}{{\dddot{k}}_{1}}-2{{\dot{k}}_{1}}){a^{2}}\arctan
r+3{{\dot{k}}_{1}}a\arctan r+({{\ddot{k}}_{2}}+4{k_{2}}){a^{2}}+3{{\dot{k}%
}_{1}}r-6{k_{2}}a
\end{equation}
Here we have grouped terms so that each term is a coefficient independent of
$r$ multiplied by a function of $r$ and the functions of $r$ are linearly
independent. Thus each coefficient must vanish, which yields ${k_{2}}=0$ and
${{\dot{k}}_{1}}=0,$ from which it follows that ${k_{1}}={c_{1}}$ and ${B^{r}%
}=0$. It then follows from Eq.(\ref{kt2}) that $A$ is independent of $z$. From
Eq.(\ref{spatialda}) it follows that
\begin{equation}
A(r)={c_{2}}+{c_{1}}{a^{-2}}%
\end{equation}
Using Eq.(\ref{decomposition}) we find that the general Killing tensor of the
$rz$ surface is
\begin{equation}
{X_{ab}}={c_{1}}{g_{ab}}+{c_{2}}{\lambda_{a}}{\lambda_{b}.}%
\end{equation}

\subsection{An $rz\phi$ surface}

We now consider an $rz\phi$ surface with metric
\begin{equation}
g_{ab}dx^{a}d{x}^{b}={a^{2}}(d{r^{2}+}d{z^{2}})+(r^{2}/a^{2})d{\phi^{2}.}%
\end{equation}
The Killing field is $\eta^{a}$, with $\epsilon=1$ and $V=r/a$. Here
$h_{\mu\nu}$ is the $g_{ab}$ of the previous subsection, while $C_{\mu\nu}$ is
the $X_{ab}$ of the previous subsection. We have
\begin{align}
{h_{\mu\nu}}  &  ={a^{2}}(r,_{\mu}r,_{\nu}+z,_{\mu}z,_{\nu})\\
{C_{\mu\nu}}  &  ={k_{1}h_{\mu\nu}}+{k_{2}}{\lambda_{\mu}\lambda}{_{\nu}.}%
\end{align}
With $B^{\mu}\rightarrow(B^{r},B^{z})$, the $rr,\,zz,$ and $rz$ components of
Eq.(\ref{kt4a}) are then, respectively
\begin{align}
{{\dot{k}}_{1}}{a^{2}}  &  =-2{\frac{r^{2}}{a^{2}}}\left(  a{\frac{da}%
{dr}B^{r}}+{a^{2}}{\partial_{r}}{B^{r}}\right) \label{ktrr}\\
{{\dot{k}}_{1}}{a^{2}}+{{\dot{k}}_{2}}{a^{4}}  &  =-2{\frac{r^{2}}{a^{2}}%
}\left(  a{\frac{da}{dr}B^{r}}+{a^{2}}{\partial_{z}}{B^{z}}\right)
\label{ktzz}\\
0  &  ={\partial_{r}}{B^{z}}+{\partial_{z}}{B^{r}} \label{ktrz}%
\end{align}
From Eq.(\ref{ktrr}) we find that $B^{r}$ must take the form
\begin{equation}
{B^{r}}={\frac{1}{a}}\left[  F(z,\phi)-\frac{{{\dot{k}}_{1}}}{2}\left(
-{\frac{1}{r}}+3r+{r^{3}}+{\frac{r^{5}}{5}}\right)  \right]
\label{brsolution}%
\end{equation}
for some function $F(z,\phi)$. Then using Eq.(\ref{brsolution}) and
integrating Eq.(\ref{ktrz}) we find
\begin{equation}
{B^{z}}=G(z,\phi)-({\frac{\partial F}{\partial z})}\text{arc}{\tan\ }r
\label{bzsolution}%
\end{equation}
for some function $G(z,\phi)$. Finally, substituting the expressions in
equations (\ref{brsolution}) and (\ref{bzsolution}) into Eq.(\ref{ktzz}) we
find
\begin{equation}
-{\frac{1}{2r^{2}}}({{\dot{k}}_{1}}{a^{2}}+{{\dot{k}}_{2}}{a^{4}}%
)={\frac{\partial G}{\partial z}}-({\frac{{\partial^{2}}F}{\partial{z^{2}}}%
)}\text{arc}{\tan\ }r+2F({\frac{r}{a^{2}})}-{\frac{{\dot{k}}_{1}}{a^{2}}%
}\left(  -1+3{r^{2}}+{r^{4}}+{\frac{r^{6}}{5}}\right)  \label{3d1}%
\end{equation}
The only odd functions of $r$ in this equation are arc${\tan\ }r$ and
$r/{a^{2}}$ and these functions are linearly independent, so the coefficient
of each must vanish. This implies that $F=0$. Furthermore, in order that the
left hand side not diverge as $r\rightarrow0$ we must have ${{\dot{k}}_{2}%
}=-{{\dot{k}}_{1}}$. Equation (\ref{3d1}) then becomes
\begin{equation}
{\frac{\partial G}{\partial z}}={{\dot{k}}_{1}}\left[  {{\frac{{a^{2}}%
(1+a)}{2}}}+{\frac{1}{a^{2}}}\left(  -1+3{r^{2}}+{r^{4}}+{\frac{r^{6}}{5}%
}\right)  \right]
\end{equation}
It then follows that both ${\dot{k}}_{1}$ and $\partial G/\partial z$ must
vanish. Thus, we have found that $C_{\mu\nu}$ and $B^{\mu}$ take the form
\begin{align}
{C_{\mu\nu}}  &  ={c_{1}}{h_{\mu\nu}}+{c_{2}}{\lambda_{\mu}\lambda}{_{\nu}}\\
{B^{\mu}}  &  ={k_{4}(\phi)}{\lambda^{\mu}}%
\end{align}
for some function ${k_{4}}(\phi)$. We now impose the integrability condition
Eq.(\ref{ic1a}) which forces ${\dot{k}}_{4}$ to vanish. This implies ${k_{4}%
}={c_{4}}$ and thus ${B^{\mu}}={c_{4}\lambda^{\mu}}$. It then follows from
Eq.(\ref{kt2a}) that $A$ is independent of $\phi$. Equation (\ref{spatialda})
then becomes
\begin{equation}
{\partial_{\mu}}A={c_{1}\partial_{\mu}(r/a)^{-2}}%
\end{equation}
for which the solution is
\begin{equation}
A(r)={c_{3}}+{c_{1}}{\frac{a^{2}}{r^{2}}}%
\end{equation}
Thus the general Killing tensor of the $rz\phi$ surface is
\begin{equation}
{X_{ab}}={c_{1}}{g_{ab}}+{c_{2}\lambda}{_{a}\lambda}{_{b}}+{c_{3}\eta}%
{_{a}\eta}{_{b}}+2{c_{4}}{\lambda_{(a}\eta}{_{b)}.}%
\end{equation}

\subsection{The Melvin metric}

We are now ready to treat the full Melvin metric by adding the $\tau^{a}$
Killing field to the metric of the previous subsection. We have
\begin{align}
{h_{\mu\nu}}  &  ={a^{2}}(r,_{\mu}r,_{\nu}+z,_{\mu}z,_{\nu})+(r^{2}/a^{2}%
)\phi,_{\mu}\phi,_{\nu}\\
{C_{\mu\nu}}  &  ={k_{1}h}{_{\mu\nu}}+{k_{2}}{\lambda}_{\mu}\lambda{_{\nu}%
}+{k_{3}}{\eta_{\mu}\eta}{_{\nu}}+2{k_{4}\lambda}{_{(\mu}\eta}{_{\nu)}}%
\end{align}
The $\tau^{a}$ Killing vector has $\epsilon=-1$ and $V=a$. Equation
(\ref{kt4a}) for ${\dot{C}_{\mu\nu}}$ becomes the following:
\begin{align}
{{\dot{k}}_{1}}  &  =2{a^{2}}{\partial_{r}}{B^{r}}+2a{\frac{da}{dr}}{B^{r}%
}\label{mkt1}\\
{{\dot{k}}_{1}}+{{\dot{k}}_{2}}{a^{2}}  &  =2a{\frac{da}{dr}}{B^{r}}+2{a^{2}%
}{\partial_{z}}{B^{z}}\label{mkt2}\\
{a^{-2}}{{\dot{k}}_{1}}+{\frac{r^{2}}{a^{4}}}{{\dot{k}}_{3}}  &
=2{\partial_{\phi}}{B^{\phi}}+2\left(  {\frac{1}{r}}-{\frac{1}{a}}{\frac
{da}{dr}}\right)  {B^{r}}\label{mkt3}\\
0  &  ={\partial_{r}}{B^{z}}+{\partial_{z}}{B^{r}}\label{mkt4}\\
0  &  ={a^{2}}{\partial_{\phi}}{B^{r}}+{\frac{r^{2}}{a^{2}}}{\partial_{r}%
}{B^{\phi}}\label{mkt5}\\
{{\dot{k}}_{4}}  &  ={\frac{a^{4}}{r^{2}}}{\partial_{\phi}}{B^{z}}%
+{\partial_{z}}{B^{\phi}} \label{mkt6}%
\end{align}
Solving Eq.(\ref{mkt1}) we find
\begin{equation}
{B^{r}}={\frac{1}{a}[}\frac{{{\dot{k}}_{1}}}{2}\text{arc}{\tan\ }%
r+F(z,\phi,t)] \label{4dBr}%
\end{equation}
for some function $F(z,\phi,t)$. Now, using this result in Eq.(\ref{mkt4}) we
find
\begin{equation}
{B^{z}}=-{\frac{\partial F}{\partial z}}\text{arc}{\tan\ }r+G(z,\phi,t)
\label{4dBz}%
\end{equation}
for some function $G(z,\phi,t)$. Using equations (\ref{4dBr}) and (\ref{4dBz})
in Eq.(\ref{mkt2}), we obtain
\begin{equation}
0=-{{\dot{k}}_{1}}+\left(  2{\frac{\partial G}{\partial z}}-{{\dot{k}}_{2}%
}\right)  {a^{2}}+4Fr+2{{\dot{k}}_{1}}r \, \text{arc}{\tan\ }r-2({\frac
{{\partial^{2}}F}{\partial{z^{2}}})}{a^{2}}\text{arc}{\tan\ }r
\end{equation}
Here we have grouped terms so that each term consists of a function of $r$
multiplied by a coefficient that is independent of $r$. Since the functions of
$r$ are linearly independent, it follows that each coefficient vanishes. We
then find that $F=0$, and that ${k_{1}}={c_{1}}$, and that $G={{\dot{k}}_{2}%
}z/2+h(\phi,t)$ for some function $h(\phi,t)$. That is, $B^{r}$ vanishes, and
$B^{z}$ takes the form
\begin{equation}
{B^{z}}={{\frac{1}{2}}}{{\dot{k}}_{2}}z+h(\phi,t)
\end{equation}
Equations (\ref{mkt3},\ref{mkt5},\ref{mkt6}) then become
\begin{align}
({\frac{r^{2}}{a^{4}})}{{\dot{k}}_{3}}  &  =2{\partial_{\phi}}{B^{\phi}%
}\label{phia}\\
0  &  ={\partial_{r}}{B^{\phi}}\label{phib}\\
{{\dot{k}}_{4}}  &  =({\frac{a^{4}}{r^{2}})}{\frac{\partial h}{\partial\phi}%
}+{\partial_{z}}{B^{\phi}} \label{phic}%
\end{align}
Differentiating equations (\ref{phia}) and (\ref{phic}) with respect to $r$,
and using Eq.(\ref{phib}), it follows that ${{\dot{k}}_{3}}=0$ and $\partial
h/\partial\phi=0$. Thus we have ${k_{3}}={c_{3}}$ for constant $c_{3}$, and
$h={k_{5}}$ for some function ${k_{5}}(t)$, and ${B^{\phi}}={{\dot{k}}_{4}%
}z+{k_{6}}$ for function ${k_{6}}(t)$. $B^{\mu}$ takes the form
\begin{equation}
{B^{\mu}}=\left(  {{\frac{1}{2}}}{{\dot{k}}_{2}}z+{k_{5}}\right)
\lambda{^{\mu}}+\left(  {{\dot{k}}_{4}}z+{k_{6}}\right)  {\eta^{\mu}}
\label{fullB}%
\end{equation}
We therefore have
\begin{equation}
{a^{-2}}{{\dot{B}}_{\mu}}=\left(  {{\frac{1}{2}}}{{\ddot{k}}_{2}}z+{{\dot{k}%
}_{5}}\right)  {\partial_{\mu}}z+{\frac{r^{2}}{a^{4}}}\left(  {{\ddot{k}}_{4}%
}z+{{\dot{k}}_{6}}\right)  {\partial_{\mu}}\phi\label{bdot}%
\end{equation}
Using the expression in Eq.(\ref{bdot}), we find that Eq.(\ref{ic1a}) becomes
\begin{equation}
0={{\ddot{k}}_{4}}z+{{\dot{k}}_{6}} \label{k-d2dot}%
\end{equation}
From Eq.(\ref{k-d2dot}) we find that ${k_{6}}={c_{6}}$ and ${k_{4}}={c_{4}%
}t+{c_{7}}$ for constants ${c_{4}},\,{c_{6}}$, and $c_{7}$. Using the
expression for $B^{\mu}$ of Eq.(\ref{fullB}) in Eq.(\ref{ic2a}), we find
\begin{equation}
0={a^{2}}\left(  {{\frac{1}{2}}}{{\dddot{k}}_{2}}z+{{\ddot{k}}_{5}}\right)
{\partial_{\mu}}z+{a^{-1}(}{\frac{da}{dr})}{{\dot{k}}_{1}}{\partial_{\mu}}r
\end{equation}
from which it follows that ${\dot{k}}_{1}$ and ${\ddot{k}}_{5}$ and
${\dddot{k}}_{2}$ all vanish. Thus we have ${k_{1}}={c_{1}}$ and ${k_{5}%
}={c_{5}}t+{c_{8}}$ and ${k_{2}}={c_{2}}{t^{2}}+{c_{9}}t+{c_{10}}$ for
constants ${c_{1}},\,{c_{2}},\,{c_{5}},\,{c_{8}},\,{c_{9},}$ and $c_{10}$.

To summarize, we have found that the general solution for $C_{\mu\nu}$ and
$B_{\mu}$ takes the form
\begin{align}
{C_{\mu\nu}}  &  ={c_{1}}{h_{\mu\nu}}+({c_{2}}{t^{2}}+{c_{9}}t+{c_{10}%
}){\lambda_{\mu}\lambda}{_{\nu}}+{c_{3}\eta}{_{\mu}\eta}{_{\nu}}+({c_{4}%
}t+{c_{7}})2\lambda{_{(\mu}\eta}{_{\nu)}}\label{finalC}\\
{B_{\mu}}  &  =\left[  \left(  {c_{2}}t+{{\frac{1}{2}}}{c_{9}}\right)
z+({c_{5}}t+{c_{8}})\right]  {\lambda_{\mu}}+({c_{4}}z+{c_{6}})\eta{_{\mu}}
\label{finalB}%
\end{align}

It remains to find $A$. Since ${B^{r}}=0$, it follows from Eq.(\ref{kt2a})
that ${\dot{A}}=0$. Using the expressions of equations (\ref{finalC}) and
(\ref{finalB}) in Eq.(\ref{spatialda}) for grad A, we find
\begin{equation}
{\partial_{\mu}}A=2({c_{2}}z+{c_{5}}){\partial_{\mu}}z-{c_{1}}{\partial_{\mu}%
}({a^{-2}}). \label{gradA}%
\end{equation}
The general solution of Eq.(\ref{gradA}) is
\begin{equation}
A={c_{2}}{z^{2}}+2{c_{5}}z-{c_{1}}{a^{-2}}+{c_{11}} \label{finalA}%
\end{equation}
Finally, using the expressions of equations (\ref{finalC}-\ref{finalA}) in
Eq.(\ref{decomposition}) we find that the most general Killing tensor of
Melvin's magnetic universe takes the 4-dimensional form
\begin{align}
{X_{ab}}  &  ={c_{1}}{g_{ab}}+{c_{2}[z^{2}}{\tau_{a}\tau}{_{b}}+zt{\lambda
_{(a}}{\tau_{b)}}+{t^{2}\lambda}{_{a}\lambda}{_{b}]}+{c_{3}\eta}{_{a}\eta
}{_{b}}\nonumber\\
&  +{c_{4}[}2z\eta{_{(a}\tau}{_{b)}}+2t{\eta_{(a}\lambda}{_{b)}]}+{c_{5}%
[}2z\tau{_{a}\tau}{_{b}}+2t{\lambda_{(a}\tau}{_{b)}]}+{c_{6}}2\eta{_{(a}\tau
}{_{b)}}\nonumber\\
&  +{c_{7}}2{\lambda_{(a}\eta}{_{b)}}+{c_{8}}2\lambda{_{(a}\tau}{_{b)}}%
+{c_{9}[}z\lambda{_{(a}\tau}{_{b)}}+t\lambda{_{a}\lambda}{_{b}]}%
+{c_{10}\lambda}{_{a}\lambda}{_{b}}+{c_{11}\tau}{_{a}\tau}{_{b}}%
\end{align}
This expression can be simplified by noting that the Melvin boost Killing
vector is given by ${\psi^{a}:}=t{\lambda^{a}}+z\tau{^{a}}$. The Killing
tensor is then%
\begin{align}
{X_{ab}}  &  ={c_{1}}{g_{ab}}+{c_{2}}{\psi_{a}}{\psi_{b}}+{c_{3}\eta}{_{a}%
\eta}{_{b}}+2{c_{4}}\eta{_{(a}}{\psi_{b)}}+2{c_{5}\psi_{(a}\tau}{_{b)}%
}+2{c_{6}\eta_{(a}\tau}{_{b)}}\nonumber\\
&  +2{c_{7}}\lambda{_{(a}\eta}{_{b)}}+2{c_{8}}\lambda{_{(a}\tau}{_{b)}}%
+{c_{9}}{\psi_{(a}\lambda}{_{b)}}+{c_{10}\lambda}{_{a}\lambda}{_{b}}%
+{c_{11}\tau}{_{a}\tau}{_{b}}%
\end{align}
Now each term in the sum is either the metric or the symmetrized product of
two Killing vectors. Thus all the Killing tensors of Melvin's magnetic
universe are trivial.

\section{Killing tensors of the Schwarzschild metric}

The Schwarzschild vacuum solution is given by
\begin{equation}
ds_{\text{Sch}}^{2}=-F dt^{2}+ {F^{-1}} dr^{2}+r^{2}(d\vartheta^{2}+\sin
^{2}\vartheta d\varphi^{2}).
\end{equation}
where the function $F$ is $F = 1 - (2m/r)$. The metric admits four Killing
vectors ($\tau^{a},\alpha^{a},\beta^{a},\gamma^{a}$); a timelike Killing
vector $\tau^{a}\partial_{a}=\partial_{t}$, and three spacelike vectors which
comprise the $SO_{3}$ rotations
\begin{align*}
\alpha^{a}\partial_{a}  &  =\sin\varphi\ \partial_{\vartheta}+\cot
\vartheta\cos\varphi\ \partial_{\varphi}\\
\beta^{a}\partial_{a}  &  =-\cos\varphi\ \partial_{\vartheta}+\cot
\vartheta\sin\varphi\ \partial_{\varphi}\\
\gamma^{a}\partial_{a}  &  =\partial_{\varphi}\\
2r^{2}  &  =\alpha^{a}\alpha_{a}+\beta^{a}\beta_{a}+\gamma^{a}\gamma_{a}.
\end{align*}
Each of the four Killing vectors is hypersurface orthogonal.

First we will find the Killing tensors of the 2-dimensional $r\vartheta$
surface. As before, we will use ${c_{1}},\,{c_{2}}$ etc. to denote constants,
and ${k_{1}},\,{k_{2}}$ etc. to denote quantities that depend only on the
coordinate associated with the Killing vector.

\subsection{An $r\vartheta$ surface}

The $r\vartheta$ 2-surface has metric%
\begin{equation}
g_{ab}dx^{a}dx^{b}=F^{-1}dr^{2}+r^{2}d\vartheta^{2} \label{r-theta-met}%
\end{equation}
We have the $\vartheta$ coordinate Killing vector $\partial_{\vartheta
}=\vartheta^{a}\partial_{a}$ for which $\epsilon=1$ and $V=r$ (note that
$\vartheta^{a}$ is a symmetry of the 2-surface but not of the spacetime). The
metric on the space orthogonal to the Killing vector is $h_{\mu\nu}%
=F^{-1}r,_{\mu}r,_{\nu}$. Since the r-direction is a 1-dimensional line, it
follows as before that the Killing tensor on that 1-dimensional space takes
the form
\begin{equation}
{C_{\mu\nu}}={k_{1}h_{\mu\nu}=k}_{1}F^{-1}r,_{\mu}r,_{\nu} \label{C-mu-nu}%
\end{equation}
for some $k_{1}(\vartheta)$. Equation (\ref{kt4a}) then becomes, with $B^{\mu
}\rightarrow B^{r}$ only
\begin{equation}
{{\dot{C}}_{rr}}={{\dot{k}}_{1}F}^{-1}=-r^{2}\left[  -(\frac{2m}{r^{2}}%
){F}^{-2}B^{r}+2{F}^{-1}\frac{dB^{r}}{dr\ }\right]
\end{equation}
or%
\begin{equation}
{{\dot{k}}_{1}}=2m{F^{-1}}B^{r}-2r^{2}\frac{dB^{r}}{dr\ }%
\end{equation}
The general solution for $B^{r}$ is
\begin{equation}
B^{r}=-(\frac{{{\dot{k}}_{1}}}{2m})F+k_{2}F^{1/2} \label{Br-soln}%
\end{equation}
Equation (\ref{ic1a}) is automatically satisfied, while Eq.(\ref{ic2a})
becomes%
\begin{equation}
\left(  {{\ddot{k}}_{2}}+{{\frac{1}{2}}}{k_{2}}\right)  {F^{-1/2}}+\left(
{\frac{{\dot{k}}_{1}}{m}}-{\frac{{\dddot{k}}_{1}}{2m}}\right)  +{{\frac{1}{2}%
}}{k_{2}}{F^{1/2}}-{\frac{3}{2m}}{{\dot{k}}_{1}}F
\end{equation}
Here we have grouped terms so that each term is a coefficient independent of
$r$ multiplied by a function of $r$, and the functions of $r$ are linearly
independent. Thus each coefficient must vanish, which yields ${k_{2}}=0$ and
${\dot{k}}_{1}=0$, from which it follows that ${k_{1}}={c_{1}}$ and ${B^{r}%
}=0$. It then follows from Eq.(\ref{kt2}) that $A$ is independent of
$\vartheta$. From Eq.(\ref{spatialda}) integration provides
\begin{equation}
A(r)=\frac{c_{1}}{r^{2}}+c_{2}.
\end{equation}
Using Eq.(\ref{decomposition}) we find that the general Killing tensor of the
$r\vartheta$ surface is
\begin{equation}
{X_{ab}}=c_{1}g_{ab}+{c_{2}\vartheta}{_{a}\vartheta}{_{b}.} \label{r-theta-X}%
\end{equation}

\subsection{A $tr\vartheta$ surface}

We now consider a $tr\vartheta$ surface with metric
\begin{equation}
g_{ab}dx^{a}d{x}^{b}=-Fd{t^{2}}+F^{-1}dr^{2}+r^{2}d\vartheta^{2}.
\end{equation}
The Killing field is $\tau^{a}\partial_{a}=\partial_{t}$, with $\epsilon=-1$
and $V={F^{1/2}}$. Here $h_{\mu\nu}$ is the $g_{ab}$ of the previous
subsection, while $C_{\mu\nu}$ is the $X_{ab}$ of the previous subsection. We
have
\begin{align}
{h_{\mu\nu}}  &  =F^{-1}r,_{\mu}r,_{\nu}+{r^{2}}\vartheta,_{\mu}%
\vartheta,_{\nu}\\
{C_{\mu\nu}}  &  ={k}_{1}h_{\mu\nu}+{k_{2}\vartheta_{\mu}\vartheta_{\nu}.}
\label{cmunu3a}%
\end{align}
With $B^{\mu}\rightarrow(B^{r},B^{\vartheta})$, the $rr,\vartheta\vartheta,$
and $r\vartheta$ components of Eq.(\ref{kt4a}) are then, respectively%
\begin{align}
{{\dot{k}}_{1}}{F^{-1}}  &  =2{\partial_{r}}{B^{r}}-{\frac{2m}{r^{2}}}{F^{-1}%
}{B^{r}}\label{sch3a}\\
{{\dot{k}}_{1}}{r^{2}}+{{\dot{k}}_{2}}{r^{4}}  &  =2{r^{2}}F{\partial
_{\vartheta}}{B^{\vartheta}}+2rF{B^{r}}\label{sch3b}\\
0  &  ={r^{2}}{\partial_{r}}{B^{\vartheta}}+{F^{-1}}{\partial_{\vartheta}%
}{B^{r}} \label{sch3c}%
\end{align}
From Eq.(\ref{sch3a}) we find that $B^{r}$ must take the form
\begin{equation}
{B^{r}}={F^{1/2}}H(t,\vartheta)+{{\frac{1}{2}}}{{\dot{k}}_{1}}\left\{
[r-6m]+3m{F^{1/2}}\ln\left[  {\frac{r}{m}}(1+{F^{1/2}})-1\right]  \right\}
\label{br1}%
\end{equation}
for some function $H(t,\vartheta)$. Using the expression of Eq.(\ref{br1}) in
Eq.(\ref{sch3c}) we find
\begin{equation}
{\partial_{r}}{B^{\vartheta}}=-{r^{-2}}{F^{-1/2}}{\partial_{\vartheta}}H.
\end{equation}
Upon integration, it follows that $B^{\vartheta}$ is
\begin{equation}
{B^{\vartheta}}=Q(t,\vartheta)-{\frac{1}{m}}{F^{1/2}}{\partial_{\vartheta}}H
\label{bth1}%
\end{equation}
for integration function $Q(t,\vartheta)$. Substituting the expressions for
${B^{r}}$ and ${B^{\vartheta}}$ from equations (\ref{br1}) and (\ref{bth1})
into Eq.(\ref{sch3b}) provides
\begin{align}
{{\dot{k}}_{1}}+{{\dot{k}}_{2}}{r^{2}}  &  ={\frac{2F}{r}}\left[  {F^{1/2}%
}H(t,\vartheta)+{{\frac{1}{2}}}{{\dot{k}}_{1}}\left\{  [r-6m]+3m{F^{1/2}}%
\ln\left[  {\frac{r}{m}}(1+{F^{1/2}})-1\right]  \right\}  \right] \nonumber\\
&  +2F\left[  {\partial_{\vartheta}}Q-{\frac{1}{m}}{F^{1/2}\frac{{\partial
^{2}}H}{\partial{\vartheta^{2}}}}\right]  . \label{sch3b2}%
\end{align}
Note that each term in Eq.(\ref{sch3b2}) is a function of $r$ multiplied by a
coefficient that is independent of $r$. Since the function of $r$ that has the
logarithmic term in Eq.(\ref{sch3b2}) is linearly independent of the other
functions of $r$, its coefficient must vanish. This implies that ${{\dot{k}%
}_{1}}=0$ and that ${k_{1}}={c_{1}}$ for some constant $c_{1}$. Equation
(\ref{sch3b2}) then simplifies to
\begin{equation}
0=-{{\frac{1}{2}}}{{\dot{k}}_{2}(}\frac{r^{2}}{F}{)}+H(\frac{{F^{1/2}}}%
{r})+{\partial_{\vartheta}}Q-{\frac{1}{m}}{\frac{{\partial^{2}}H}%
{\partial{\vartheta^{2}}}(}{F^{1/2}).} \label{sch3b3}%
\end{equation}
Here terms are grouped so that each term is a coefficient independent of $r$
multiplied by a function of $r$, and so that the functions of $r$ are linearly
independent. It therefore follows that each of the coefficients vanishes. We
have ${{\dot{k}}_{2}}=0$, $H=0$, and ${\partial_{\vartheta}}Q=0$. Therefore
${k_{2}}={c_{2}}$ for some constant $c_{2}$, and the components of vector
field $B^{\mu}$ are
\begin{equation}
{B^{r}}=0,\text{ \ }{B^{\vartheta}}={k_{3}}(t)
\end{equation}
for some function ${k_{3}}(t)$. Equivalently
\begin{equation}
{B_{\mu}}={k_{3}}{r^{2}}{\partial_{\mu}}\vartheta. \label{bmu3a}%
\end{equation}
Since ${k_{1}}={c_{1}}$ and ${k_{2}}={c_{2},}$ it follows from
Eq.(\ref{cmunu3a}) that tensor $C_{\mu\nu}$ takes the form
\begin{equation}
{C_{\mu\nu}}={c_{1}}{h_{\mu\nu}}+{c}_{2}{r^{4}}{\vartheta_{,\mu}}%
{\vartheta_{,\nu}} \label{cmunu3b}%
\end{equation}
Upon using equations (\ref{bmu3a}) and (\ref{cmunu3b}) in Eq.(\ref{ic1a}) we
find that
\begin{equation}
{\partial_{\lbrack\mu}}({F^{-1}}{r^{2}}{{\dot{k}}_{3}}{\partial_{\nu]}%
}\vartheta)=0
\end{equation}
from which it follows that ${{\dot{k}}_{3}}=0$ and therefore that ${k_{3}%
}={c_{3}}$ for some constant $c_{3}$. Thus, from Eq.(\ref{bmu3a}) we have
\begin{equation}
{B_{\mu}}={c_{3}}{r^{2}}{\partial_{\mu}}\vartheta. \label{bmu3b}%
\end{equation}
Equation (\ref{ic2a}) is identically satisfied by equations (\ref{cmunu3b})
and (\ref{bmu3b}).

Since ${B^{r}}=0$ it follows from Eq.(\ref{kt2a}) that ${\dot{A}}=0$. Using
equations (\ref{cmunu3b}) and (\ref{bmu3b}) in Eq.(\ref{spatialda}) we obtain
\begin{equation}
{\partial_{\mu}}A=-{c_{1}}{\partial_{\mu}}({F^{-1}}).
\end{equation}
It then follows that
\begin{equation}
A={c_{4}}-{c_{1}}{F^{-1}} \label{a3}%
\end{equation}
for some constant $c_{4}$. Finally, using the results of equations
(\ref{cmunu3b}), (\ref{bmu3b}) and (\ref{a3}) in Eq.(\ref{decomposition}), we
find that the general Killing tensor of a $tr\theta$ surface is
\begin{equation}
{X_{ab}}={c_{1}}{g_{ab}}+{c_{2}}{\vartheta_{a}}{\vartheta_{b}}+2{c_{3}%
}{\vartheta_{(a}}{\tau_{b)}}+{c_{4}}{\tau_{a}}{\tau_{b}}%
\end{equation}

\subsection{The Schwarzschild metric}

We are now ready to treat the full Schwarzschild metric by adding the axial
$\gamma^{a}$ Killing vector to the metric of the previous subsection. We have
\begin{align}
{h_{\mu\nu}}  &  =-F{t_{,\mu}}{t_{,\nu}}+{F^{-1}}{r_{,\mu}}{r_{,\nu}}+{r^{2}%
}{\vartheta_{,\mu}}{\vartheta_{,\nu}}\\
{C_{\mu\nu}}  &  ={k_{1}}{h_{\mu\nu}}+{k_{2}}{\vartheta_{\mu}}{\vartheta_{\nu
}}+2{k_{3}}{\vartheta_{(\mu}}{\tau_{\nu)}}+{k_{4}}{\tau_{\mu}}{\tau_{\nu}}%
\end{align}
The $\gamma^{a}$ Killing vector has $\epsilon=1$ and $V=r\sin\vartheta$.
Equation (\ref{kt4a}) for ${\dot{C}}_{\mu\nu}$ then becomes
\begin{align}
-{{\dot{k}}_{1}}F+{{\dot{k}}_{4}}{F^{2}}  &  ={r^{2}}{\sin^{2}}\vartheta
({B^{r}}{\partial_{r}}F+2F{\partial_{t}}{B^{t}})\label{scha}\\
{{\dot{k}}_{1}}{F^{-1}}  &  ={r^{2}}{\sin^{2}}\vartheta({F^{-2}}{B^{r}%
}{\partial_{r}}F-2{F^{-1}}{\partial_{r}}{B^{r}})\label{schb}\\
{{\dot{k}}_{1}}{r^{2}}+{{\dot{k}}_{2}}{r^{4}}  &  =-2{r^{3}}{\sin^{2}%
}\vartheta({B^{r}}+r{\partial_{\vartheta}}{B^{\vartheta}})\label{schc}\\
0  &  ={F^{-1}}{\partial_{t}}{B^{r}}-F{\partial_{r}}{B^{t}}\label{schd}\\
-{{\dot{k}}_{3}}{r^{2}}F  &  ={r^{2}}{\sin^{2}}\vartheta(F{\partial
_{\vartheta}}{B^{t}}-{r^{2}}{\partial_{t}}{B^{\vartheta}})\label{sche}\\
0  &  ={r^{2}}{\partial_{r}}{B^{\vartheta}}+{F^{-1}}{\partial_{\vartheta}%
}{B^{r}} \label{schf}%
\end{align}
Equation (\ref{schb}) can be rewritten as
\begin{equation}
{\partial_{r}}({F^{-1/2}}{B^{r}})=-{\frac{{\dot{k}}_{1}}{2F^{1/2}{r^{2}}%
{\sin^{2}}\vartheta}}%
\end{equation}
with integral
\begin{equation}
{B^{r}}=G(t,\vartheta,\varphi){F^{1/2}}-{\frac{{\dot{k}}_{1}}{2m{\sin^{2}%
}\vartheta}}F \label{bra}%
\end{equation}
for integration function $G(t,\vartheta,\varphi)$. Using the result of
Eq.(\ref{bra}) in Eq.(\ref{schf}), we find
\begin{equation}
{\partial_{r}}{B^{\vartheta}}=-\frac{{\partial_{\vartheta}}G}{r^{2}F^{1/2}%
}-{\frac{{{\dot{k}}_{1}}\cos\vartheta}{mr^{2}{\sin^{3}}\vartheta}.}%
\end{equation}
Integration provides
\begin{equation}
{B^{\vartheta}}=H(t,\vartheta,\varphi)-({\partial_{\vartheta}}G){\frac{1}{m}%
}{F^{1/2}}+{\frac{{{\dot{k}}_{1}}\cos\vartheta}{mr{\sin^{3}}\vartheta}}
\label{btha}%
\end{equation}
with integration function $H(t,\vartheta,\varphi)$. Now using equations
(\ref{bra}) and (\ref{btha}) in Eq.(\ref{schc}) we obtain
\begin{equation}
0=-3{{\dot{k}}_{1}}-({{\dot{k}}_{2}}+{\partial_{\vartheta}}H\,2{\sin^{2}%
}\vartheta){r^{2}}+{\frac{{\dot{k}}_{1}}{m}}({\frac{6}{{\sin^{2}}\vartheta}%
}-3)r-2{\sin^{2}}\vartheta G(r{F^{1/2})}+{\frac{2}{m}}{\sin^{2}}%
\vartheta({\frac{{\partial^{2}}G}{\partial{\vartheta^{2}}})}{r^{2}}{F^{1/2}}%
\end{equation}
Here we have grouped terms so that each term is a coefficient independent of
$r$ multiplied by a function of $r,$ and so that the functions of $r$ are
linearly independent. It therefore follows that each coefficient vanishes.
Thus $G={\dot{k}}_{1}=0$ and
\begin{equation}
{\partial_{\vartheta}}H=-{\frac{{\dot{k}}_{2}}{2{\sin^{2}}\vartheta}.}
\label{dhdth}%
\end{equation}
From the vanishing of $G$ and ${\dot{k}}_{1}$ it follows that $B^{r}$
vanishes, and that ${k_{1}}={c_{1}}$ for some constant $c_{1}$ and that
${B^{\vartheta}}=H$. From Eq.(\ref{dhdth}) it follows that ${B^{\vartheta}}$
takes the form
\begin{equation}
{B^{\vartheta}}=I(t,\varphi)+{\frac{{\dot{k}}_{2}}{2}}\cot\vartheta
\label{bthb}%
\end{equation}
for some function $I(t,\varphi)$. Using Eq.(\ref{bthb}), along with $B^{r}=0$
and ${\dot{k}}_{1}=0,$ reduces equations (\ref{scha}-\ref{schf}) to the
following:
\begin{align}
{\partial_{t}}{B^{t}}  &  ={\frac{{{\dot{k}}_{4}}F}{2{r^{2}}{\sin^{2}%
}\vartheta}}\label{dtbt}\\
{\partial_{r}}{B^{t}}  &  =0\label{drbt}\\
{\partial_{\vartheta}}{B^{t}}  &  ={r^{2}}{F^{-1}}{\partial_{t}}I-{\frac
{{\dot{k}}_{3}}{{\sin^{2}}\vartheta}} \label{dthbt}%
\end{align}
Applying $\partial_{r}$ to Eq.(\ref{dtbt}) and using Eq.(\ref{drbt}) yields
${\dot{k}}_{4}=0$. Therefore $B^{t}$ is independent of $t$ and ${k_{4}}%
={c_{2}}$ for some constant $c_{2}$. Now applying $\partial_{r}$ to
Eq.(\ref{dthbt}) and using Eq.(\ref{drbt}) it follows that ${\partial_{t}}I=0$
and $I={k_{5}}$ for some function ${k_{5}}(\varphi)$. Thus $B^{\vartheta}$
becomes
\begin{equation}
{B^{\vartheta}}={k_{5}}+{\frac{{\dot{k}}_{2}}{2}}\cot\vartheta.
\end{equation}
Integrating Eq.(\ref{dthbt}) yields
\begin{equation}
{B^{t}}={k_{6}}+{{\dot{k}}_{3}}\cot\vartheta
\end{equation}
for some function ${k_{6}}(\varphi)$. In summary, we have found that $B_{\mu}$
and $C_{\mu\nu}$ take the form
\begin{align}
{B_{\mu}}  &  =({k_{6}}+{{\dot{k}}_{3}}\cot\vartheta){\tau_{\mu}}+({k_{5}%
}+{\frac{{\dot{k}}_{2}}{2}}\cot\vartheta){\vartheta_{\mu}}\label{bmua}\\
{C_{\mu\nu}}  &  ={c_{1}}{h_{\mu\nu}}+{k_{2}}{\vartheta_{\mu}}{\vartheta_{\nu
}}+2{k_{3}}{\vartheta_{(\mu}}{\tau_{\nu)}}+{c_{2}}{\tau_{\mu}}{\tau_{\nu}}
\label{cmunua}%
\end{align}

We now impose the integrability conditions of equations (\ref{ic1a}) and
(\ref{ic2a}) on the expressions above for $B_{\mu}$ and $C_{\mu\nu}$. From
equations (\ref{bmua}) and (\ref{cmunua}), with $\epsilon=1$ and
$V=r\sin\vartheta$, we find
\begin{align}
{V^{-2}}{{\dot{B}}_{\mu}}+{V^{-3}}{C_{\mu\nu}}{\partial^{\nu}}V  &  ={c_{1}%
}{V^{-3}}{\partial_{\mu}}V-{\frac{F}{r^{2}{\sin^{2}}\vartheta}}\left[
{{\dot{k}}_{6}}+({{\ddot{k}}_{3}}+{k_{3}})\cot\vartheta\right]  {\partial
_{\mu}}t\nonumber\\
&  +{\frac{1}{{\sin^{2}}\vartheta}}\left[  {{\dot{k}}_{5}}+({\frac{{\ddot{k}%
}_{2}}{2}}+{k_{2})}\cot\vartheta\right]  {\partial_{\mu}}\vartheta
\label{curlfree}%
\end{align}
The integrability condition given in Eq.(\ref{ic1a}) is the statement that the
right hand side of Eq.(\ref{curlfree}) is curl-free. From this it follows
that, for the term in Eq.(\ref{curlfree}) multiplying ${\partial_{\mu}}t$, the
quantity in square brackets vanishes. That is, we have ${{\dot{k}}_{6}}=0$ and
${{\ddot{k}}_{3}}+{k_{3}}=0$. Thus
\begin{align}
{k_{6}}  &  ={c_{3}}\label{k6}\\
{k_{3}}  &  ={c_{4}}\cos\varphi+{c_{5}}\sin\varphi\label{k3}%
\end{align}
for constants ${c_{3}},\,{c_{4}}$, and $c_{5}$.

From equations (\ref{bmua}), (\ref{curlfree}), (\ref{k6}), and (\ref{k3}) it
follows that
\begin{equation}
{{\ddot{B}}_{\mu}}+{V^{-1}}{{\dot{C}}_{\mu\nu}}{\partial^{\nu}} V
-\epsilon{V^{2}}{\partial_{\mu}}({V^{-1}}{B^{\nu}}{\partial_{\nu}}V)=\left[
{{\ddot{k}}_{5}}+{k_{5}}+({\frac{{\dddot{k}}_{2}}{2}}+2{{\dot{k}}_{2})}%
\cot\vartheta\right]  {r^{2}}{\partial_{\mu}}\vartheta\label{schic2}%
\end{equation}
The integrability condition of Eq.(\ref{ic2a}) states that the right hand side
of Eq.(\ref{schic2}) vanishes. Therefore ${{\ddot{k}}_{5}}+{k_{5}}=0$ and
${{\dddot{k}}_{2}}+4{{\dot{k}}_{2}}=0$, and thus
\begin{align}
{k_{5}}  &  ={c_{6}}\cos\varphi+{c_{7}}\sin\varphi\\
{k_{2}}  &  ={c_{8}}\cos2\varphi+{c_{9}}\sin2\varphi+{c_{10}}%
\end{align}
for constants ${c_{6}},\,{c_{7}},\,{c_{8}},\,{c_{9}},$ and $c_{10}$. We find
that the general solution for $B_{\mu}$ and $C_{\mu\nu}$ is
\begin{align}
{B_{\mu}}  &  =[{c_{3}}+(-{c_{4}}\sin\varphi+{c_{5}}\cos\varphi)\cot
\vartheta]{\tau_{\mu}}\nonumber\\
&  +[{c_{6}}\cos\varphi+{c_{7}}\sin\varphi+(-{c_{8}}\sin2\varphi+{c_{9}}%
\cos2\varphi)\cot\vartheta]{\vartheta_{\mu}}\label{bmub}\\
{C_{\mu\nu}}  &  ={c_{1}}{h_{\mu\nu}}+({c_{8}}\cos2\varphi+{c_{9}}\sin
2\varphi+{c_{10}}){\vartheta_{\mu}}{\vartheta_{\nu}}\nonumber\\
&  +2({c_{4}}\cos\varphi+{c_{5}}\sin\varphi){\vartheta_{(\mu}}{\tau_{\nu)}%
}+{c_{2}}{\tau_{\mu}}{\tau_{\nu}} \label{cmunub}%
\end{align}

It remains to find $A$. Using equations (\ref{bmub}) and (\ref{cmunub}) in
Eq.(\ref{spatialda}) we obtain
\begin{align}
{\partial_{\mu}}A  &  =-2{c_{1}}{V^{-3}}{\partial_{\mu}}V\\
&  -{\frac{2}{{\sin^{2}}\vartheta}}[-{c_{6}}\sin\varphi+{c_{7}}\cos
\varphi+(-{c_{8}}\cos2\varphi-{c_{9}}\sin2\varphi+{c_{10}})\cot\vartheta
]{\partial_{\mu}}\vartheta,\nonumber
\end{align}
with general solution
\begin{equation}
A={c_{1}}{V^{-2}}+2(-{c_{6}}\sin\varphi+{c_{7}}\cos\varphi)\cot\vartheta
+(-{c_{8}}\cos2\varphi-{c_{9}}\sin2\varphi+{c_{10}}){\cot^{2}}\vartheta
+{k_{7}} \label{asolsch1}%
\end{equation}
for some function ${k_{7}}(\varphi)$. Imposing Eq.(\ref{kt2a}) on
Eq.(\ref{asolsch1}) we find that ${{\dot{k}}_{7}}=0$ and therefore that
${k_{7}}={c_{11}}$ for some constant $c_{11}$. The solution for $A$ becomes
\begin{equation}
A={c_{1}}{V^{-2}}+2(-{c_{6}}\sin\varphi+{c_{7}}\cos\varphi)\cot\vartheta
+(-{c_{8}}\cos2\varphi-{c_{9}}\sin2\varphi+{c_{10}}){\cot^{2}}\vartheta
+{c_{11}} \label{asolsch2}%
\end{equation}
Using equations (\ref{bmub}), (\ref{cmunub}), and (\ref{asolsch2}) in
Eq.(\ref{decomposition}), and grouping terms according to their constant
coefficient, we find that the general Schwarzschild Killing tensor is
\begin{align}
{X_{ab}}  &  ={c_{1}}({h_{ab}}+{V^{-2}}{\gamma_{a}}{\gamma_{b}})+{c_{2}}%
{\tau_{a}}{\tau_{b}}+{c_{3}}2{\tau_{(a}}{\gamma_{b)}}\nonumber\\
&  +{c_{4}[}2\cos\varphi{\vartheta_{(a}}{\tau_{b)}}-2\sin\varphi\cot
\vartheta{\gamma_{(a}}{\tau_{b)}]}\nonumber\\
&  +{c_{5}[}2\sin\varphi{\vartheta_{(a}}{\tau_{b)}}+2\cos\varphi\cot
\vartheta{\gamma_{(a}}{\tau_{b)}]}+{c_{6}[}2\cos\varphi{\vartheta_{(a}}%
{\gamma_{b)}}-2\sin\varphi\cot\vartheta{\gamma_{a}}{\gamma_{b}]}\nonumber\\
&  +{c_{7}[}2\sin\varphi{\vartheta_{(a}}{\gamma_{b)}}+2\cos\varphi
\cot\vartheta{\gamma_{a}}{\gamma_{b}]}\nonumber\\
&  +{c_{8}[}\cos2\varphi{\vartheta_{a}}{\vartheta_{b}}-2\sin2\varphi
\cot\vartheta{\vartheta_{(a}}{\gamma_{b)}}-\cos2\varphi{\cot^{2}}%
\vartheta{\gamma_{a}}{\gamma_{b}]}\nonumber\\
&  +{c_{9}[}\sin2\varphi{\vartheta_{a}}{\vartheta_{b}}+2\cos2\varphi
\cot\vartheta{\vartheta_{(a}}{\gamma_{b)}}-\sin2\varphi{\cot^{2}}%
\vartheta{\gamma_{a}}{\gamma_{b}]}\nonumber\\
&  +{c_{10}[\vartheta_{a}}{\vartheta_{b}}+{\cot^{2}}\vartheta{\gamma_{a}%
}{\gamma_{b}]}+{c_{11}}{\gamma_{a}}{\gamma_{b}}%
\end{align}
We now rewrite this expression for the Killing tensor in terms of Killing
vectors and the metric. We have
\begin{align}
{g_{ab}}  &  ={h_{ab}}+{V^{-2}}{\gamma_{a}}{\gamma_{b},}\\
{\alpha_{a}}  &  =\sin\varphi\,{\vartheta_{a}}+(\cot\vartheta\cos
\varphi){\gamma_{a},}\\
{\beta_{a}}  &  =-\cos\varphi\,{\vartheta_{a}}+(\cot\vartheta\sin
\varphi){\gamma_{a},}%
\end{align}
which yields
\begin{align}
{\alpha_{a}}{\alpha_{b}}+{\beta_{a}}{\beta_{b}}  &  ={\vartheta_{a}}%
{\vartheta_{b}}+{\cot^{2}}\vartheta{\gamma_{a}}{\gamma_{b}}\\
{\alpha_{a}}{\alpha_{b}}-{\beta_{a}}{\beta_{b}}  &  =\cos2\varphi
(-{\vartheta_{a}}{\vartheta_{b}}+{\cot^{2}}\vartheta{\gamma_{a}}{\gamma_{b}%
})+2\cot\vartheta\sin2\varphi{\vartheta_{(a}}{\gamma_{b)}}\\
2{\alpha_{(a}}{\beta_{b)}}  &  =\sin2\varphi(-{\vartheta_{a}}{\vartheta_{b}%
}+{\cot^{2}}\vartheta{\gamma_{a}}{\gamma_{b}})-2\cot\vartheta\cos
2\varphi{\vartheta_{(a}}{\gamma_{b)}}%
\end{align}
Using the three equations above, we can rewrite $X_{ab}$ as
\begin{align}
{X_{ab}}  &  ={c_{1}}{g_{ab}}+{c_{2}}{\tau_{a}}{\tau_{b}}+{c_{3}}2{\tau_{(a}%
}{\gamma_{b)}}-2{c_{4}}{\beta_{(a}}{\tau_{b)}}+2{c_{5}}{\alpha_{(a}}{\tau
_{b)}}-2{c_{6}}{\beta_{(a}}{\gamma_{b)}}+2{c_{7}}{\alpha_{(a}}{\gamma_{b)}%
}\nonumber\\
&  -{c_{8}}({\alpha_{a}}{\alpha_{b}}-{\beta_{a}}{\beta_{b}})-2{c_{9}}%
{\alpha_{(a}}{\beta_{b)}}+{c_{10}}({\alpha_{a}}{\alpha_{b}}+{\beta_{a}}%
{\beta_{b}})+{c_{11}}{\gamma_{a}}{\gamma_{b}.}%
\end{align}
Finally, regrouping terms we have
\begin{align}
{X_{ab}}  &  ={c_{1}}{g_{ab}}+{c_{2}}{\tau_{a}}{\tau_{b}}+{c_{3}}2{\tau_{(a}%
}{\gamma_{b)}}-2{c_{4}}{\beta_{(a}}{\tau_{b)}}+2{c_{5}}{\alpha_{(a}}{\tau
_{b)}}-2{c_{6}}{\beta_{(a}}{\gamma_{b)}}+2{c_{7}}{\alpha_{(a}}{\gamma_{b)}%
}\nonumber\\
&  +({c_{10}}-{c_{8}}){\alpha_{a}}{\alpha_{b}}+({c_{10}}+{c_{8}}){\beta_{a}%
}{\beta_{b}}-2{c_{9}}{\alpha_{(a}}{\beta_{b)}}+{c_{11}}{\gamma_{a}}{\gamma
_{b}.}%
\end{align}
Thus, we have written the general Schwarzschild Killing tensor as a sum of
terms where each term is either the metric or a product of Killing vectors.
Therefore all Killing tensors of the Schwarzschild spacetime are trivial.

\section{Conclusions}

The method used above consists of applying equations {(\ref{kt4a}%
-\ref{spatialda})} to find the Killing tensor in n-dimensions, using the
equations in n-1 dimensions. For the Melvin metric and the Schwarzschild
metric this is done three times, going from a 1-dimensional space to the
4-dimensional spacetime.

Professor G. Valent has noted that his recent paper \cite{Val04} uses a
similar method for studying geodesic flows. His method differs from ours.
Valent assumes the Killing tensor is Lie derived by the Killing field. We make
no such assumption. Here, we assume the Killing field is hypersurface orthogonal.

The method developed here for finding Killing tensors could be used on a wide
variety of spacetimes where there are symmetries. It requires less computation
than an attack on the full Killing tensor equations, and it can be used even
when the spacetime is not algebraically special. The method could also be
generalized in various ways. The equations for a Killing-Yano tensor
\cite{Col76} could be treated in an analogous way and should result in a
method simpler than a straightforward attempt to solve the Killing-Yano
equations. Finally the method might have a useful generalization to the case
where the Killing vector is not hypersurface orthogonal. In that case one
would expect to get more complicated equations that involve not only the norm
of the Killing field but also the twist. However, it is in just such
spacetimes (i.e. Kerr) that known examples of nontrivial Killing tensors
exist. So an investigation along those lines might be useful.

\end{document}